# Exploring assessment method of technological advancement based on literature cross-citation

Shengxuan Tang, Liming Zhang, Shuo Jiang, Ming Cai, Yao Xiao*

School of Intelligent Systems Engineering, Sun Yat-Sen University, Shenzhen, Guangdong, China

*Abstract*—Assessing advancements of technology is essential for creating science and technology policies and making informed investments in the technology market. However, current methods primarily focus on the characteristics of the technologies themselves, making it difficult to accurately assess technologies across various fields and generations. To address this challenge, we propose a novel approach that uses bibliometrics, specifically literature citation networks, to measure changes in knowledge flow throughout the evolution of technology. This method can identify diverse trends in technology development and is an effective tool for evaluating technological advancements. We demonstrate its accuracy and applicability by applying it to mobile communication technology and comparing its quantitative results with other assessment methods. Our work provides critical support for assessing different technical routes and formulating technology policy.

## 1. Introduction

Science and technology are widely recognized as crucial drivers of social and economic progress. Accurately assessing the advancements and relative advantages of different technologies can help determine optimal technological pathways, establish rational resource allocation strategies, and identify key research priorities. As such, evaluating and forecasting various technologies, understanding technological progress and trends, and identifying dynamic areas of technological development is essential.

Technology assessment has long been a major concern in various fields and is one of the core research directions in technology management. The main methods currently used for technology assessment include the Delphi method [1-4], scenario analysis [5-8], and decision analysis [9-12]. The Delphi method is a questionnaire-based method that organizes and shares opinions through feedback. It has four distinct characteristics: anonymity, iterative nature, feedback, and statistical "group response". Kareem A [2] developed a fuzzy Delphi method consisting of three steps: usability criteria analysis, fuzzy Delphi analysis, and usability evaluation model development. Majed G [3] proposed a variant of the Delphi method using triangular fuzzy numbers with a similar communication method to experts but a different evaluation process. Tomas Eloy Salais-Fierro [4] established a hybrid approach combining expert judgment with demand forecasting generated from historical data for use in the automotive industry.

Scenario analysis is a widely used method for technology assessment in many management fields and is particularly valuable for assessing technology. This approach generally considers the impacts of multiple scenarios on technologies. Guo [5] et al. designed four scenarios with different policy considerations - a baseline scenario, a subsidy scenario, a low carbon price scenario, and a high carbon price scenario - to evaluate and select energy-efficient technology routes across the supply chain. Hussain [6] constructed five scenarios using end-user and econometric approaches. Gómez-Sanabria [7] et al. included key governance variables in their analysis of different scenarios to assess the potential impacts

---

* Corresponding author.

 *E-mail address:* xiaoyao9@mail.sysu.edu.cn (Y. Xiao).

of greenhouse gases from seven wastewater treatment options in the Indonesian fish processing industry. To make scenarios more realistic, Gelder [8] et al. proposed a scenario representativeness (SR) metric based on Wasserstein distance that quantifies how well generated parameter values represent real scenarios while covering actual changes found in real scenarios.

Decision analysis methods have also been applied to technology assessments. The decision analysis approach for technology assessment typically involves two steps: developing a decision analysis framework and ranking the performance of each technology in the study area using the constructed framework. Liang [9] et al. developed a fuzzy group decision support framework for prioritizing the sustainability of alternative fuel-based vehicles. Dahooie et al [10] used a fuzzy multi-attribute decision-making (F-MADM) method to rank interactive television technologies. To enhance the credibility of decision analysis results, other tools can be incorporated into the decision analysis framework. Zeng et al [11] proposed a fuzzy group decision support framework and introduced a Pythagorean fuzzy aggregation operator to compensate for existing shortcomings and enhance credibility. They demonstrated the complete process of their multi-criteria decision-making (MCDM) model by evaluating unmanned ground transportation technology and conducting comparative and sensitivity analyses. Dahooie et al [12] proposed an integrated framework based on sentiment analysis (SA) and MCDM technology using intuitionistic fuzzy sets (IFS). They evaluated and ranked five cell phone products using online customer reviews (OCR) on Amazon.com to illustrate the usability and usefulness of OCR.

Through a review of the three technology assessment methods (i.e., the Delphi method, scenario analysis, and decision analysis), it is clear that these methods have been applied in various fields. However, they still have some inherent limitations. The Delphi method lacks communication of ideas and may be subject to subjective bias and a tendency to ignore minority opinions. This can lead to assessment results that deviate from reality and are influenced by the organizer's subjectivity. Scenario analysis is time-consuming and costly and relies on assumptions that are still in flux, which can affect the reliability of results. Although this method has an objective basis, it relies heavily on the analyst's assumptions about data and is therefore influenced by their value orientation and subjectivity. The decision analysis method has a limited scope of use and cannot be applied to some decisions that cannot be expressed quantitatively. The determination of the probability of occurrence of various options is sometimes more subjective and may lead to poor decisions.

All the three methods, the Delphi method, scenario analysis, and decision analysis, belong to technology mining methods. Technology mining methods assess technology from the perspective of the technology itself, while data mining methods use data related to the technology for assessment. The key difference between these two approaches is that technology mining is problem-oriented and past studies have focused on the technology itself rather than directly searching for patterns in data. In this work, we introduce technology-related data for use in technology assessment analysis.

To overcome the aforementioned limitations, a concept of technological knowledge flows(TKFs) can be utilized because the citation information on scientific papers or patent has been considered a reliable proxy for TKF [13-15]. Park et al [13] proposes a future-oriented approach to discovering technological opportunities for convergence using patent information and a link prediction method in a directed network. Chen et al [14] tests the hypothesis that patent citations indicate knowledge linkage by measuring text similarities between citing-cited patent pairs and finds that examiner citations are a better indicator of knowledge linkage than applicant citations. Żogała-Siudem et al [15] introduces a reparametrized versions of the discrete generalized beta distribution and power law models that preserve the total sum of elements in a citation vector, resulting in better predictive power and easier numerical fitting.

Fundamental science research is a crucial form of scientific activity and the foundation of technological progress. Most breakthrough technological achievements benefit from fundamental science research, which embodies knowledge flow. New technologies generally build on existing ones and are constantly innovated and accumulated until disruptive innovation occurs at a certain node. This represents another manifestation of technological knowledge flow and transmission. Fundamental science research produces theorized concepts and essences as its main outputs and typically uses journal literature as a carrier for its knowledge. Since a research field is not simply composed of journal literature, its overall knowledge architecture network is constructed through cross-citations among journal literature. These cross-citation relationships can reflect technological development to some extent. In this paper, we explore technology advancement assessment from the perspective of knowledge flow by analyzing literature data and their cross-citation relationships.

Section 2 introduces the sources and the processing methods of data, and a technology assessment model is formulated. The mobile communication technology (i.e., 2G – 6G) is applied as an example in Section 3 to analyze the availability of the assessment model. In Section 4, other technology assessment methods are introduced to compare the performance with our cross-citation-based assessment model. Based on the assessment results, Section 5 provides a summary of the assessment model and analyzes the potential limitations and improvements.

## 2. Data and Models

The research methodology employed in this study encompasses two main components: (1) Data collection and pre-processing, and (2) Cross-citation calculation and model formulation. The first component involves retrieving potentially relevant literature from the Web of Science core database using key search terms related to the technical domain of interest and refining the collected literature using a binary classifier that distinguishes between relevant and irrelevant documents based on term frequency. The second component entails calculating cross-citations between corresponding documents for each technology in the domain to generate a cross-citation matrix, and subsequently proposing a technology assessment model that incorporates the cross-citation matrix to obtain advanced assessment results for each technology.

.

### 2.1. Data Collection and Pre-processing

Web of Science is a citation database maintained by the Institute for Scientific Information (ISI) that contains over 8,000 highly influential, peer-reviewed journals from around the world. For this paper, all research data was sourced from the Web of Science database due to its comprehensiveness and reliability. The way to obtain literature data in this paper is by keyword search. All literature retrieved from the database was exported as txt files for further data processing. However, the literature obtained by keyword search alone will have a lot of interfering data, which is because many articles simply mention that the keyword will also be retrieved or some keywords have multiple layers of meaning, and the literature with all the meanings involved in the keyword will be retrieved by keyword search. The authenticity of the data has a great impact on the data processing results and also on the final conclusions. Therefore, it is a critical task to filter the target literature from the literature obtained by keyword search. Here, we improve a support vector machine (SVM) approach [16] to perform the relevant/irrelevant dichotomous classification of the literature retrieved through keywords:

**Step 1: Text Preprocessing**. In this stage, the abstract section of the literature is exclusively extracted and subjected to a series of preprocessing steps such as division, deactivation, and normalization. This process aims to enhance the quality and readability of the text.

**Step 2: Feature Extraction**. The text information is transformed into numerical vectors using the bag-of-words model, Term Frequency-Inverse Document Frequency (TF-IDF), and other techniques. These numerical vectors serve as input features for the SVM model.

**Step 3: Model Training**. Researchers manually annotate a subset of documents to facilitate model training. In the second round of testing, randomly selected annotated documents ensure the accuracy of the labeled literature dataset. The ratio of relevant literature to irrelevant literature is 1:1, and 80% of the literature is selected randomly for training, while the remaining 20% is reserved for classification effect evaluation. The SVM model is trained using the labeled literature dataset to determine the hyperplane with the maximum interval on the feature space for separating the relevant and irrelevant literature. The kernel function is chosen as the linear kernel function, with a penalty factor of C=1.0.

**Step 4: Model Evaluation**. The performance of the SVM model in terms of classification and generalization abilities on unknown data is evaluated using cross-validation, confusion matrix, accuracy, recall, and other relevant metrics.

**Step 5: Document Classification**. The trained SVM text classification model is applied to all the literature, and documents judged as relevant are labeled with "1," while those deemed irrelevant are labeled with "0." Only documents labeled "1" are retained for later cross-citation calculation.

The training sample size in Step 3 also has a great impact on the model training, so the sample size for SVM classifier learning is determined before the parameters are determined. In this paper, the SVM classifier is trained using different number of sample sizes for each technique separately. The resulting classifiers are then used to classify the entire literature corresponding to each technique separately. The selection of the sample size is determined by the stability of the classification situation, and the formula for quantifying the stability of the classification effect is shown in Eq.1.

$$\sigma_n = |\omega_i(n) - \omega_i(n-100)|/P_i, \quad n > 100 \tag{1}$$

Where $\sigma_n$ denotes the degree of classification stability of literature using the SVM classifier. $\omega_i(n)$ notes the number of target literature obtained after classification using SVM classifier when the training sample size is n. $P_i$ denotes the initial amount of literature corresponding to technology $t_i$.

In this paper, we assume that the classification effect is sufficiently stable and meets our requirements when $\sigma_n$ is less than 0.01.

## 2.2. Cross-citation Calculation and Model Formulation

In the previous step, we preprocessed the target literature for cross-citation analysis. In order to get the cross-citation data between the literature corresponding to the two technologies, we developed a Python program to calculate the cross-citation between technology $a$ and technology $b$. First, we extracted the DOI of technology $a$ corresponding to each literature and put them all into the set $M$. Then we extracted the DOI of technology $b$ corresponding to all references in each literature and put them all into the set $N$. The citation relationship between technology $a$ and technology $b$ is shown in Eq. 2. $d_{ij}$ indicates the citation relationship between the two technologies. $d_{ij} = 1$ means that the literature of technology b cites the literature of technology a. On the contrary, $d_{ij} = 0$ means no citation relationship, which is because the literature and DOI are one-to-one correspondence.

Set $M$, and Set $N$

$$d_{ij} = \begin{cases} 1, & \text{if } M[i] = N[j] \\ 0, & \text{otherwise} \end{cases} \tag{2}$$

According to the citation relationship between technology $a$ and technology $b$ shown in Eq. 2. The citation number $c_i^j$ of technology $b$ from technology $a$ is calculated as follow.

$$C_i^j = \Sigma_{m=1}^{|M|} \Sigma_{n=1}^{|N|} d_{ij} \tag{3}$$

The same can be obtained from the number of technology $a$ citing technology $b$. Then we can calculate the number of two cross-citation between several technologies in a certain technical field according to the formula for cross-citation between two technologies given in Eq. 3. After calculating the cross-citations for all technologies, we obtained a matrix of technology cross-citations as shown in Table 1. In this matrix, $t_i$ represents technology $i$ and $C_i^j$ represents the number of citations from technology $i$ to technology $j$. Here's how we use this matrix to analyze technological advancement.

**Table 1**
Cross-citation matrix of literature corresponding to technology.

|       | $t_1$   | $t_2$   | ... | $t_i$   | ... | $t_j$   |
|-------|---------|---------|-----|---------|-----|---------|
| $t_1$ | -       | $C_1^2$ | ... | $C_1^i$ | ... | $C_1^j$ |
| $t_2$ | $C_2^1$ | -       | ... | $C_2^i$ | ... | $C_2^j$ |
| ...   | ...     | ...     | -   | ...     | ... | ...     |
| $t_i$ | $C_i^1$ | $C_i^2$ | ... | -       | ... | $C_i^j$ |
| ...   | ...     | ...     | ... | ...     | -   | ...     |
| $t_j$ | $C_j^1$ | $C_j^2$ | ... | $C_j^i$ | ... | -       |

Scientific publications are a key way to characterize scientific and technological progress and are an essential part of the human knowledge system. Previous publications serve as important references for subsequent research, allowing knowledge to accumulate over time. Citations in scientific publications show which previous literature contributed to a paper and provide context for knowledge accumulation [17-18]. Citations have been widely used in bibliometric studies to assess technological development, research performance, and map knowledge evolution or technological trajectories. Cross-citation relationships between scientific literature can reflect technological convergence and diffusion effects. Technological convergence is reflected through backward citations (references), while technological diffusion is reflected through forward citations (cited literature). The degree of entry and exit in the knowledge network constructed through citation relationships corresponds to the integration and diffusion of knowledge [19].

In-degree index $r(i)$ refers to the number of times literature $i$ cites other literature. A higher in-degree indicates that literature $i$ cites more other literature, representing a greater momentum of knowledge convergence from other sources to literature $i$

$$r(i) = \sum_{j=1, j \neq i}^{k} C_i^j \tag{4}$$

Where $k$ denotes number of technologies.

Out-degree index $c(i)$ refers to the number of times literature $i$ is cited by other literature. A higher out-degree indicates that literature $i$ is cited more by other literature, representing a greater momentum of knowledge diffusion from literature i to other sources.

$$c(i) = \sum_{j=1, j \neq i}^{k} C_j^i \tag{5}$$

**Hypothesis 1** For two types of technology $i$ and $j$, based on cross-citation data, if $C_i^j$ is greater than $C_j^i$, it suggests that technology $t_i$ may be more advanced than technology $t_j$.

$$Z_i > Z_j, \quad \text{if } C_i^j > C_j^i, \tag{6}$$

where $Z_i$, $Z_j$ denotes the advancement value of technology $i$ and $j$.

Inspired by the physical nature of knowledge fusion and diffusion reflected in the out-degree of citation networks, if technology $i$ cites technology $j$ more often than technology $j$ cites technology $i$, then technology $i$ is considered more advanced than technology $j$. The reason is considered as that technology $i$ incorporates relevant achievements from technology $j$ to make a technological breakthrough and ultimately form a new and more advanced technology.

For two technologies, the technology assessment results can be determined using their ratio. However, for three or more technologies, the technology assessment results cannot be quantified by the ratio of cross-citations. Instead, the cross-citations among them form a matrix of literature cross-citations as shown in Table 1. To take advantage of the comprehensive information of other technologies, the technology advancement index of technology $Z_i$ is formulated as follow.

$$Z_i = \frac{1}{k-1} \cdot \sum_{i=1, i \neq j}^{k} \left( \frac{C_i^j + 1}{C_j^i + 1} \cdot \log_a(C_i^j C_j^i + b) \right) / \sum_{i=1, i \neq j}^{k} \log_a(C_i^j C_j^i + b), \qquad k \geq 2 \qquad (7)$$

The scale effect of technology is considered in the work by calculating the article number of related technologies, whereas the effect of self-citation is ignored. Note that the value of parameter $a$ does not affect the final results of technology advancement

**Proof 1**.

$$Z_i = \frac{1}{k-1} \cdot \sum_{i=1, i \neq j}^{k} \left( \frac{C_i^j + 1}{C_j^i + 1} \cdot \log_a(C_i^j C_j^i + b) \right) / \sum_{i=1, i \neq j}^{k} \log_a(C_i^j C_j^i + b)$$

$$= \frac{1}{k-1} \cdot \sum_{i=1, i \neq j}^{k} \left( \frac{C_i^j + 1}{C_j^i + 1} \cdot \ln(C_i^j C_j^i + b) / \ln a \right) / \sum_{i=1, i \neq j}^{k} \ln(C_i^j C_j^i + b) / \ln a$$

$$= \frac{1}{k-1} \cdot \sum_{i=1, i \neq j}^{k} \left( \frac{C_i^j + 1}{C_j^i + 1} \cdot \ln(C_i^j C_j^i + b) \right) / \sum_{i=1, i \neq j}^{k} \ln(C_i^j C_j^i + b)$$

As a result, $Z_i$ remains the same regardless of the value of $a$.

Parameter $b$ is used to prevent negative scale effects when the reference between technologies is 0. For ease of calculation, both $a$ and $b$ are set to 2 in the formula.

## 3. Result Analysis

In this paper, we analyze the field of mobile communication technology for technology assessment. We chose this field for two main reasons. First, mobile communication technology has been a significant achievement in the development of electronic computers and the mobile internet. Over the past half-century, it has profoundly impacted various aspects of society including lifestyle, production, work, entertainment, politics, economy and culture. It is one of major technologies that have changed the world. Second, mobile communication technologies include 2G, 3G, 4G, 5G and 6G technologies. The progression of these five technologies is relatively intuitive and clear which makes it convenient to verify the technology assessment method presented in this paper.

For this paper, we selected data from published literature in the category of mobile communication technologies. The data was sourced from the Web of Science core database. The keywords used in the search were "2G", "3G", "4G", "5G" and "6G"; the types of literature selected were papers, conference proceedings, review papers, and online publications; the web of science categories selected were Engineering Electrical Electronic and Telecommunications; the publishers selected were IEEE, Elsevier and Springer Nature; the period from 2000 to 2022. The data was exported on August 30, 2022. Table 2 shows the amount of data collected using a search style for mobile communication technologies.

**Table 2**

The initial volume of literature was retrieved by search formula for mobile communication technology.

| Technology | 2G | 3G | 4G | 5G | 6G |
|---|---|---|---|---|---|
| Volume of Literature | 5152 | 7967 | 7866 | 33050 | 5281 |

Since the amount of literature for the five technologies from 2G to 6G varies significantly, this paper first determines the sample size to be used for SVM classifier learning. We trained the SVM classifier using different sample sizes for each of these five technologies. The resulting classifiers were then used to classify all literature corresponding to each technology. We calculated the variation of classification effect with sample size for literature corresponding to each technology using Eq. 1. Fig. 1 shows a schematic diagram of how the number of classified literature stabilizes with training sample size.

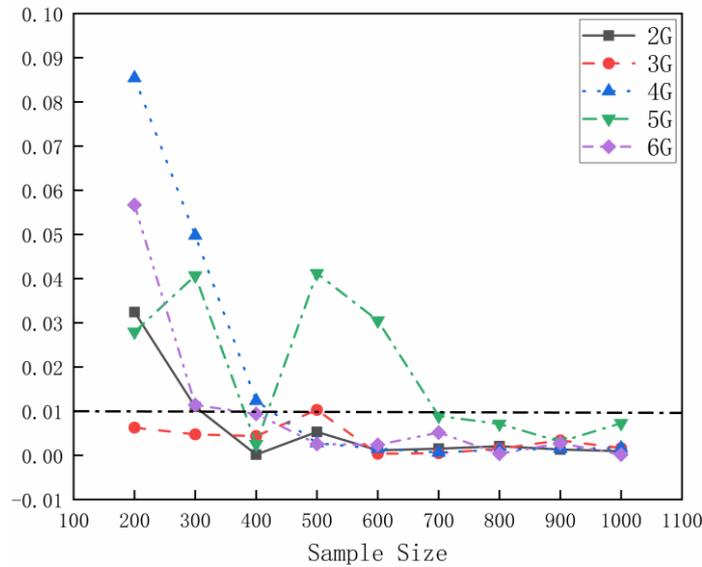

**Fig.1** The relationship between the stability of mobile communication technology classification and sample size.

From Fig. 1, we found that the classification effects of all five technologies stabilize (i.e., $\sigma_n < 0.01$) when the sample size exceeds 700 (350 relevant papers and 350 irrelevant papers). In this paper, we uniformly chose a classification sample size of 1000 for all five technologies. We conducted 10 randomized experiments for each technology and calculated the average accuracy of these classifications. Table 3 shows the average accuracy of the 10 classifications for each technology.

**Table 3**. Average of 10-test accuracy of the 5 technologies on the SVM classifier.

| Technology | 2G | 3G | 4G | 5G | 6G |
|---|---|---|---|---|---|
| Accuracy | 0.9867 | 0.9367 | 0.9960 | 0.9767 | 0.9965 |

Table 3 shows that using SVM to classify relevant and irrelevant literature for the five technologies is very effective. The classified literature can be used for subsequent cross-citation analysis.

We calculated the two-citation intercalation between each technology in the field of mobile communication technology using the Eq.3. After obtaining the two-citation intercalation between all technologies, we generated a literature inter-citation matrix for mobile communication technology as shown in Table 4.

**Table 4**

Matrix of literature cross-citations corresponding to mobile communication technologies.

|     | 2G  | 3G   | 4G    | 5G   | 6G   |
| --- | --- | ---- | ----- | ---- | ---- |
| **2G** | -   | 417  | 231   | 259  | 6    |
| **3G** | 451 | -    | 742   | 730  | 24   |
| **4G** | 403 | 1427 | -     | 4132 | 110  |
| **5G** | 739 | 2271 | 13729 | -    | 2159 |
| **6G** | 17  | 54   | 525   | 5819 | -    |

The matrix of interleaved quantities is entered from Table 4 into the technology assessment model corresponding to Eq. 7 in this paper. This allowed us to obtain quantitative results for the advancement of each mobile communication technology as shown in Fig. 2. When quantified using all years of literature, the results for technological advancement of mobile communication technologies align with our expectations in terms of size relationships. Additionally, we found that the technology assessment gap between 4G, 5G and 6G is significantly higher than that between 2G, 3G and 4G. This is because our paper quantifies relative technological sophistication and the latest mobile communication technologies (5G and 6G) are much more sophisticated than earlier generations (2G, 3G and 4G).

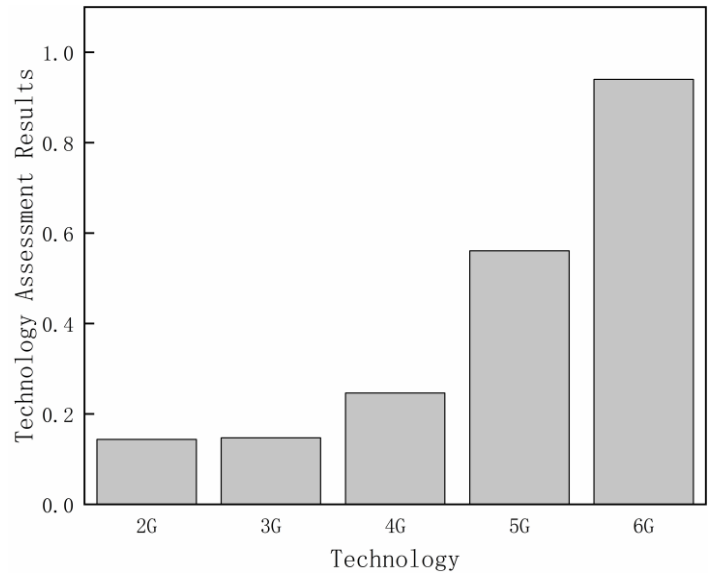

**Fig.2** Mobile Communication Technology Assessment Results.

To obtain the development trend of each technology over time, we performed a year-by-year quantification of mobile communication technology. We first counted the volume of literature on mobile communication technologies as shown in Fig. 3 (a). In order to avoid unstable technology assessment results due to too little literature data, the initial year of technology assessment was set as 2010 (i.e., the literature used for technology assessment in 2010 was the literature from 2000-2010), and the initial year of technology assessment must also satisfy that the literature volume in that year has reached 0.01 of the total literature volume of the technology. satisfying the above two conditions, the initial year of technology assessment was finally determined as 2010 for 2G, 3G, and 4G, 2014 for 5G, and 2019 for 6G.the technology assessment results of mobile communication technologies over time are calculated as shown in Fig. 3(b).

From Fig. 3 (a), we can see that the literature volume for three technologies (2G to 4G) has peaked and is now decreasing while the literature volume for 5G and 6G continues to increase year by year. the reason for the decrease in the volume of literature in 2022 is that the time for deriving the literature data is August 30, 2022, and not for a full year of literature in 2022.Currently, research interest in 5G far exceeds that of other technologies.

From Fig. 3 (b), We quantified the technology assessment of the five technologies from 2010-2021, but the technology assessment results for only 2G, 3G, and 4G in 2010 are due to the fact that the five technologies were not born at the same time; only 2G, 3G, and 4G were available in 2010, and new technologies will continue to enter as time progresses (e.g., 5G joined in 2014 and 6G joined in 2019). When an emerging technology enters the assessment, it immediately gains an advantage in technology assessment. This is because the methodology in this paper is based on cross-citation, so when a new technology first appears, it usually cites the old technology a bit more. At the same time, When an emerging technology enters the assessment, it also has some degree of impact (usually negative) on the technology assessment of already existing technologies, for example, when 5G technology joined the assessment in 2014, there was a precipitous drop in the assessment results of 4G, and a similar situation occurred in 2019, when 6G joined the assessment, there was a significant drop in the technology assessment results of 5G. The reason for this situation is that the method in this paper is a class of quantitative methods based on technology cross-referencing, and in fact we find that the impact of the entry of new technologies on different technological advancement values varies, mainly because the latest technologies always cite the most recent generation of technologies in the process of writing papers, and this behavior in our method of assessment based on literature cross-citation. This behavior also affects to some extent the technology evaluation results of the previous generation technologies. Nevertheless, the accuracy of our method in identifying the most advanced technologies is still quite high, and the highest values of technology evaluation results for each year from 2010 to 2021 correspond to the best technologies in practice.

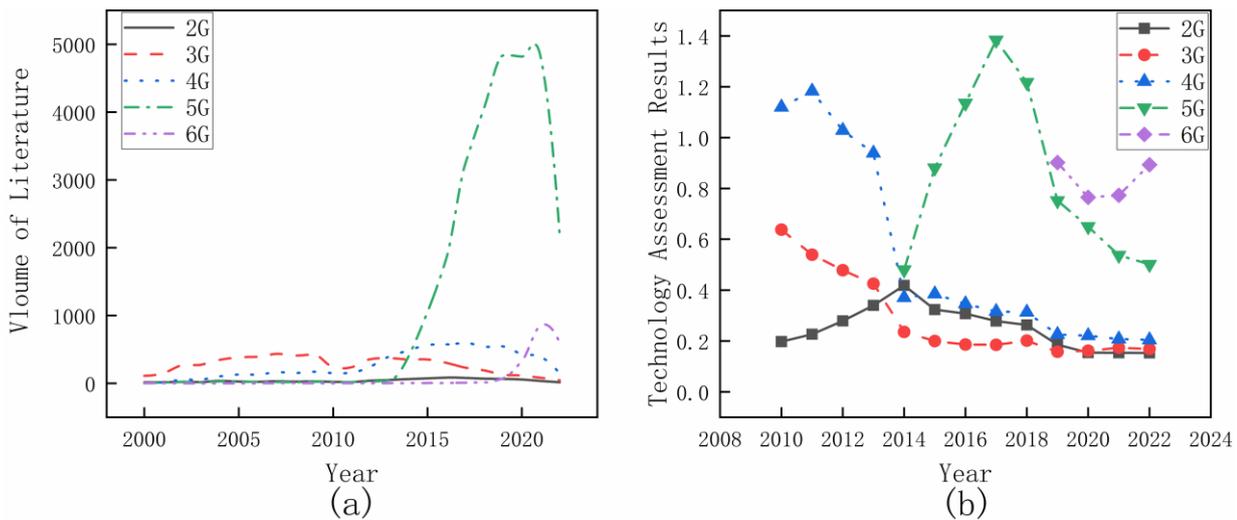

**Fig.3** Results of mobile communication technology over the years. **(a)** Literature volume **(b)** Technology assessment.

The following quantitative analysis of technological advancement using other relevant bibliometric methods further validates the reliability of the quantitative approach in this paper.

## 4. Comparative Analysis

To better analyze the characteristics of our proposed method, several technology assessment methods based on citation analysis are introduced to quantitatively analyze mobile communication technology. We compared the quantitative results to illustrate the advantages of our proposed method and its implications for technology assessment and forecasting work.

**4.1. Method Introduction**

Citation analysis has emerged as a prevalent tool in bibliometric research, serving to evaluate technological advancements, research performance, and the progression of knowledge or technological trajectories. Gutiérrez-Salcedo employed the h-index and g-index to delineate research focal points and technological trends [20]. Acosta examined the relationship between scientific and technological growth across Spain's regions by analyzing the links between science and technology through scientific citations in patent literature [21]. Hall et al. posited that, among all patent-related indicators, patent citations offer a more suitable measure for appraising market value [22]. Stuart and Podolny utilized patent citations to gauge technological progress and technology transfer within companies [23].

In knowledge networks developed via citation connections, increased centrality signifies a greater number of connections to network participants. Brass and Burkhardt contended that, from an organizational behavior standpoint, an individual with higher centrality in a social network wields greater power [24]. The network nodes directly linked to a specific node reside within that node's domain. The quantity of neighboring nodes is termed node degree or connectivity degree. Granovetter asserts that node degree is proportional to the likelihood of acquiring resources [25]. Node degree represents the extent of a node's participation in the network, serving as the foundational concept for measuring centrality. In his 2010 study, Pei-Chun Lee [20] evaluated the significance of technology by constructing a patent citation network to compute out-degree and in-degree centrality. In the present paper, we adopt Lee's methodology to assemble a complex network of literature citations encompassing five technologies (2G to 6G). Subsequently, we calculate the degree centrality for each technology to quantify it, and compare our findings with those obtained using our proposed method.

(1) **h-index**. The retrieved papers for a specific technology are ordered based on their citation frequencies. The h-index associated with the technology is defined as the value of $h$, such that the top $h$ papers each have a citation frequency of at least $h$, while the citation frequency of the $(h+1)$th paper is less than $(h+1)$. Denoting the citation frequency of the hth paper as $x_i$, the mathematical representation of the h-index can be expressed as:

$$h = max(i) : x_i \geq i \qquad (8)$$

The h-index adeptly merges two key indicators, namely the number of publications associated with a particular technology and the citation frequency reflecting the literature's quality. This approach overcomes the limitations of relying solely on a single indicator to quantify technological advancement.

(2) **g-index**. Similarly, the g-index considers the retrieved literature of technology ranked by citation frequency. The g-index is defined for a given technology such that the total number of citations garnered by the top g papers is no less than $g^2$, while the total number of citations for the $(g+1)$th paper is less than $(g+1)^2$. The mathematical representation of the g-index can be expressed as:

$$g = \max(i) : x_i \geq i^2 \qquad (9)$$

(3) **In-degree centrality**. The ratio of the number of references to other literature in literature $i$ to the total number of references to other literature in all literature. The higher the centrality of entry, the greater the momentum of knowledge integration from other literature to literature $i$. The higher the centrality of entry, the greater the momentum of knowledge integration from other literature to literature $i$.

$$r'(i) = r(i) / \sum_{i=1}^{k} r(i) \qquad (10)$$

(4) **Out-degree centrality**. The ratio of the number of citations of literature $i$ by other literature to the total number of citations of all literature. The higher the centrality, the greater the momentum of knowledge dissemination from literature $i$ to other literature.

$$c'(i) = c(i) / \sum_{i=1}^{k} c(i) \qquad (11)$$

**4.2. Results Comparison**

The outcomes of the technology evaluation utilizing the h-index and g-index are depicted in Fig. 4(a) and Fig. 4(b). To facilitate a more intuitive comparison, the h-index and g-index for each technology were transformed into percentages relative to the h-index and g-index for all five technologies.

Fig. 4(a) and Fig. 4(b) demonstrate that when employing the h-index and g-index to measure technological progress, the results are more accurate for technologies (2G to 5G) with longer development durations. However, for technologies (6G) with briefer development periods, the outcomes do not align with the actual situation due to the limited literature available. The quantification results for assessing mobile communication technology using in-degree centrality are presented in Fig. 4(c) and Fig. 4(d). It is evident from these figures that when relying solely on in-degree centrality for evaluating mobile communication technologies, the quantification results for 2G, 3G, and 4G are superior, yet there are noticeable discrepancies between the quantification outcomes for 5G and 6G in comparison to the real situation. When utilizing out-degree centrality exclusively to quantify mobile communication technologies, the quantification outcomes for each technology over time exhibit apparent size relationship inaccuracies, resulting in an overall suboptimal quantification.

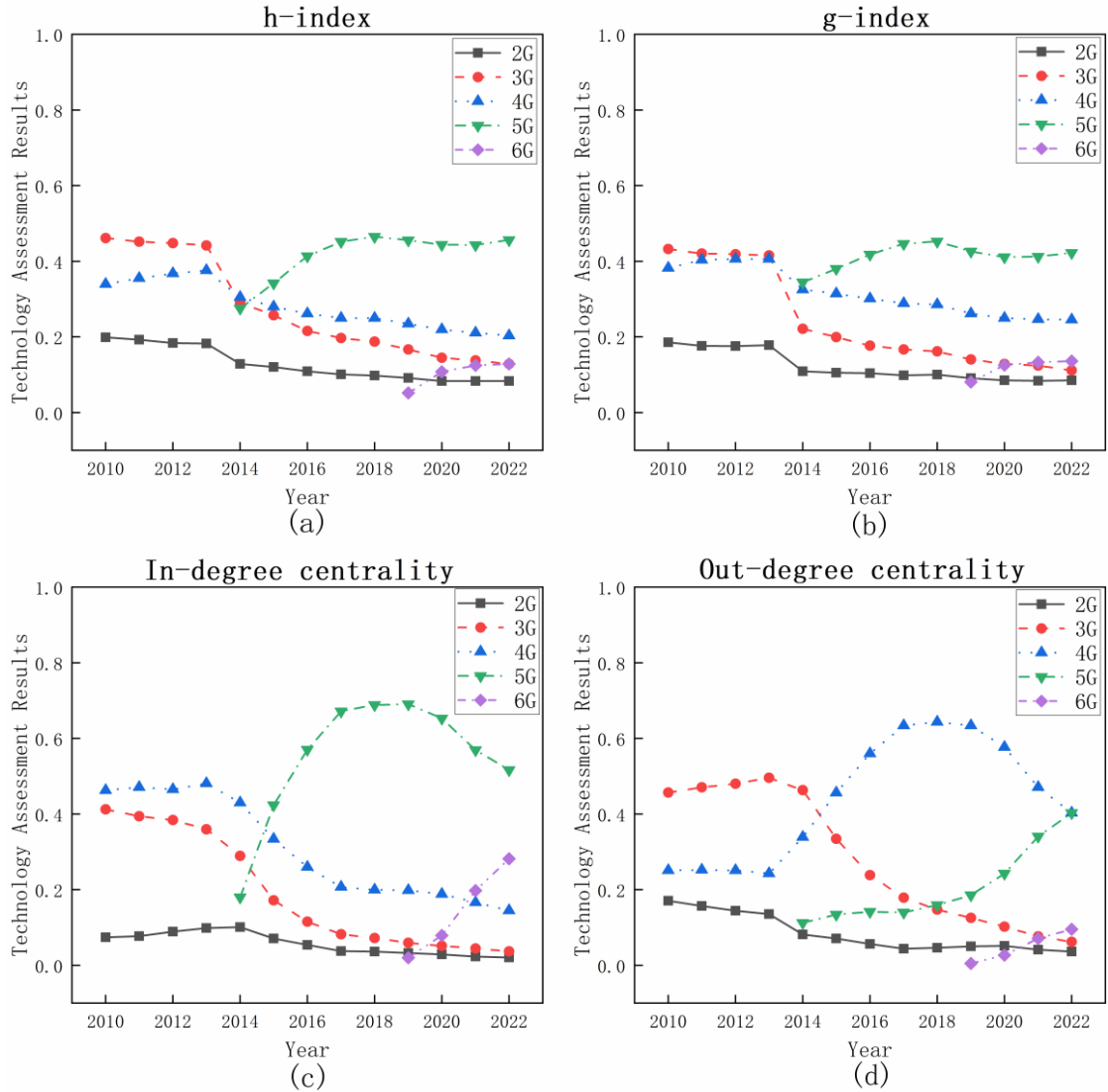

**Fig.4** Results of the technology assessment of mobile communication technologies using other methods.

The results of our quantitative analysis of literature data for each technology's full-time period are presented in Fig. 5. In addition to the methods presented in this paper, the ranking of the technology assessment of the five 2G-6G technologies in all other methods does not match the actual situation. The most obvious error is that the assessment results of the other four methods for 6G are not optimal among the five 2G-6G technologies. When using In-degree centrality for technology assessment, 4G is optimal, while the other three methods are all 5G optimal. The reason for this situation is that there are four technology assessment methods besides this paper, and the results all depend on the number of literature and citations (citation and cited quantity). The cross citation method proposed in this paper evaluates the technology from the perspective of knowledge flow.

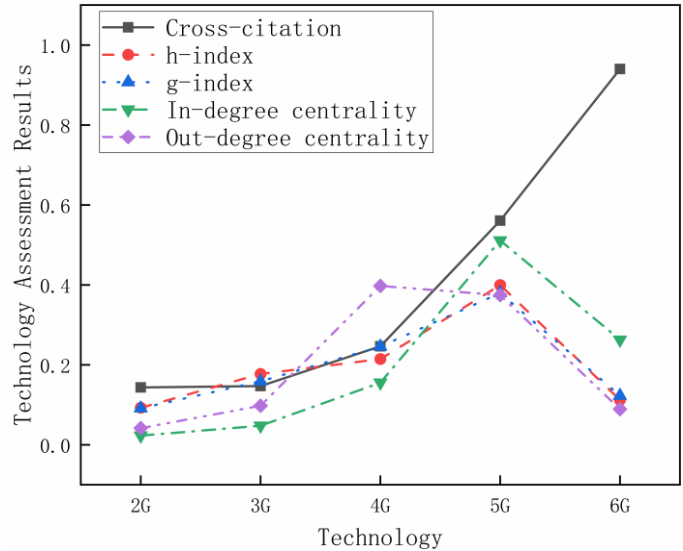

**Fig.5** Comparison of the results obtained by each technology assessment method over the full period

Fig. 6 shows the overall accuracy and mean annual accuracy of the ranking when the proposed method in this paper is used together with the other four methods for technology assessment. As shown in Figure 6, using the method proposed in this paper, both accuracy rates are at the highest level, and for the mean annual accuracy, the two methods, using our method and using In-degree centrality, are optimal. It is worth noting that when using the cross-citation based method proposed in this paper for technology assessment, the overall accuracy reached 100%, far higher than the other four methods.

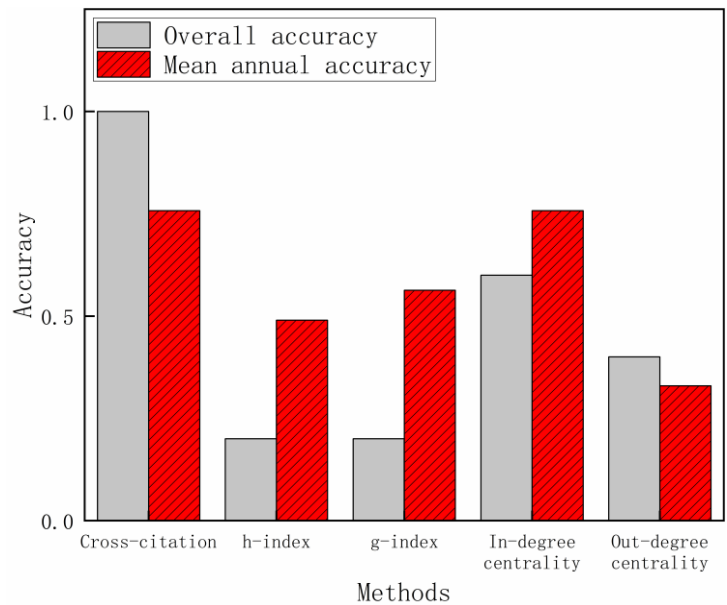

**Fig.6** The overall accuracy and mean annual accuracy of technology assessment ranking.

When using the four methods of h-index, g-index, In-degree centrality, and Out-degree centrality for technology assessment, the average level of overall accuracy and mean annual accuracy is lower than the method based on cross-citation proposed in this paper. This is due to the fact that quantifying technological advancement using the h-index and g-index requires a sufficiently long period of technological development and a large body of literature. On the other hand, quantifying technological advancement using access centrality is too one-sided; more citations to a technology's literature indicate stronger diffusion momentum while more citations to other literature indicate stronger convergence momentum. However, neither diffusion nor convergence momentum can be equated directly with technological sophistication.

Our proposed cross-citation-based method overcomes these shortcomings by integrating cited and citing data to measure technological advancement through a combination of diffusion and convergence momentum. The greatest contribution of our method is its ability to identify the optimal technology in a given field for each period. our cross-citation based method outperforms others with 100% accuracy and high stability.

## 5. Conclusion and Prospect

### 5.1. Conclusion

In this paper, we analyze the technology assessment in terms of the cross-citations among the literature corresponding to the technologies, so as to obtain the strengths and weaknesses of each technology in a series of technologies in a certain field and the future trends of the field. By comparing the h-index, g-index and access centrality methods for technology quantification, it can be seen that the method of this paper is more applicable and overcomes not only the disadvantage of using h-index and g-index for technology assessment, which requires a sufficient amount of literature accumulation but also the problem that access centrality alone is too one-sided and poorly interpretable for quantifying technological advancement.

The technology quantification method proposed in this paper, despite its many advantages, also has certain limitations.

**Data limitation**. On the one hand, the publication of literature generally appears several months later than the technology recorded on the literature record, which creates a time gap between the literature record and the technology development; on the other hand, the technology development is influenced by scientific and technological, social, economic, and policy factors, and these factors cannot all be taken into account when evaluating the technology through the method of this paper.

**Quantification formula**. The premise of quantifying technological advancement based on the amount of inter-citation is that there are at least 2 technologies, and the quantification formula in this paper cannot be quantified in the face of a single technology. In the future, we can try to integrate more factors related to technological relevance in the quantification work to optimize the quantification results of technological advancement.

**Prediction problem**. Since the data source of this paper is the scientific literature, this leads to the fact that, the technology quantification work done in this paper can only quantify the past tense, and the method cannot be implemented for the quantification results of the existing technologies at a certain period in the future.

### 5.2. Prospect

The quantification of technology through cross-citations among literature is not effective in individual cases due to the reasons mentioned above, but it is one of the influential and effective methods in technology quantification work. To overcome these limitations, other evaluation indicators that have an impact on technological advancement can be considered with cross-citation as the main indicator.

**Time-series analysis.** This method takes a chronological approach to examine the changes in the volume of various types of literature over time, fully revealing the temporal characteristics of the literature, reflecting the process of technological development and the level reached and allowing one to grasp the development status and advantages of a certain technical field.

**Hot spots of technological development.** Journals and magazines have collected various technical literature according to the principle of discipline classification or theme, and through the comparative analysis of the changes in the volume of its various types of literature or to understand the increase or decrease in the category and the reasons, from which to study the focus of technological development and the frontier of possible breakthroughs.

In addition, this paper has analyzed the change in the quantitative results of mobile communication technology with the year, and the next step can predict the technological advancement of existing technologies after 5 years and 10 years based on the technology assessment results of existing years.

# Acknowledgement

The paper was funded by the National Key R&D Program of China [grant number 2020YFB1600400].